\documentclass[aps,prl,reprint,showpacs,preprintnumbers,amsmath,amssymb,citeautoscript, superscriptaddress]{revtex4-1}

\usepackage{epsfig}
\usepackage{wasysym}
\usepackage{amssymb}
\usepackage{gensymb}  % To use \degree
\usepackage{physics}
\usepackage{color}
\usepackage{url}
\usepackage{floatrow} % To create caption on the side
\begin{document}

%\tighten
\newcommand{\bvec}[1]{\mbox{\boldmath ${#1}$}}
%\newcommand{\bm}{\boldmath}
%\preprint{\vbox{ \hfill FIS-UI-TH-XXXX}}
\title{Magnetic Tip Trap System}
\author{Oki Gunawan}\email[]{ogunawa@us.ibm.com}
\affiliation{IBM T. J. Watson Research Center, Yorktown Heights, NY 10598, USA}
\author{Jason Kristiano}
\affiliation{IBM T. J. Watson Research Center, Yorktown Heights, NY 10598, USA}
\affiliation{Simetri Foundation, Tangerang, Indonesia, 15334}
\author{Hendra Kwee}
\affiliation{Simetri Foundation, Tangerang, Indonesia, 15334}
\affiliation{Surya College of Education, Tangerang, Indonesia, 15115}
\date{\today}

\begin{abstract}
 We report a detailed theoretical model of recently-demonstrated magnetic trap system based on a pair of magnetic tips. The model takes into account key parameters such as tip diameter, facet angle and gap separation. It yields quantitative descriptions consistent with experiments such as the vertical and radial frequency, equilibrium position and the optimum facet angle that produces the strongest confinement. We arrive at striking conclusions that a maximum confinement enhancement can be achieved at an optimum facet angle $\theta_{max}=\arccos{\sqrt{2/3}}$ and a critical gap exists beyond which this enhancement effect no longer applies. This magnetic trap and its theoretical model serves as a new and interesting example of a simple and elementary magnetic trap in physics.
 \end{abstract}

\maketitle

Various electromagnetic trap systems play important role in physics for their ability to trap and isolate particles or matter that have produced many applications and discoveries.  Examples are Penning trap \cite{PenningPhys36, BrownRMP86}, optical dipole trap or optical tweezer \cite{AshkinJQE00,ChuPRL86, ChuPRL85}, magneto optic trap \cite{RaabPRL87, PhillipsRMP98} and various diamagnetic traps \cite{SimonAJP01, LyuksyutovAPL04, GunawanAPL15, HsuSR16,HoultonRSI18}. For diamagnetic trap systems, high field-gradient product ($B ~\nabla B$) is necessary to achieve trapping or levitation \cite{SimonJAP00}. A new approach is to use magnetic tip geometry as recently demonstrated by O'Brien \textit{et~al.} \cite{ObrienAPL19}. The tip geometry maximizes $B ~\nabla B$ at the trapped object that leads to stronger field confinement, thus high frequency and high quality factor ($Q$). This characteristics is of high interest for research that explores macroscopic limits of classical mechanics and quantum mechanics. Such magnetic trap also offers interesting alternative to optical trap as the latter can lead to excessive heating and  encounters instabilities in vacuum \cite{SlezakNJP18}. The ability to achieve high magnetic field gradients in a localized position using magnetic tip is also useful for other applications such as for nuclear magnetic resonant imaging \cite{MaminNatNano07} and magnetic force microscopy \cite{RugarNat04}.

Currently, there is strong interest in magnetic trap system for various applications such as precision gravimetry \cite{JohnssonSR16}, study of displacement and velocity of Brownian particle \cite{LiSci10}, gas temperature measurement \cite{MillenNatNano14}, and research that explores the boundaries of the classical and quantum systems. For example, trapped nanodiamond can be used to investigate the quantum mechanical properties such as superposition of states \cite{YinPRA13,ScalaPRL13}, control of electron spin of nanodiamond nitrogen-vacancy centers and to observe the electron spin resonance properties \cite{HoangNatComm16}. Such a trap could also serve an  important role to test quantum mechanical properties of gravity \cite{KafriNJP14,BosePRL17}.

\begin{figure}[bp]
	\includegraphics[width=\textwidth]{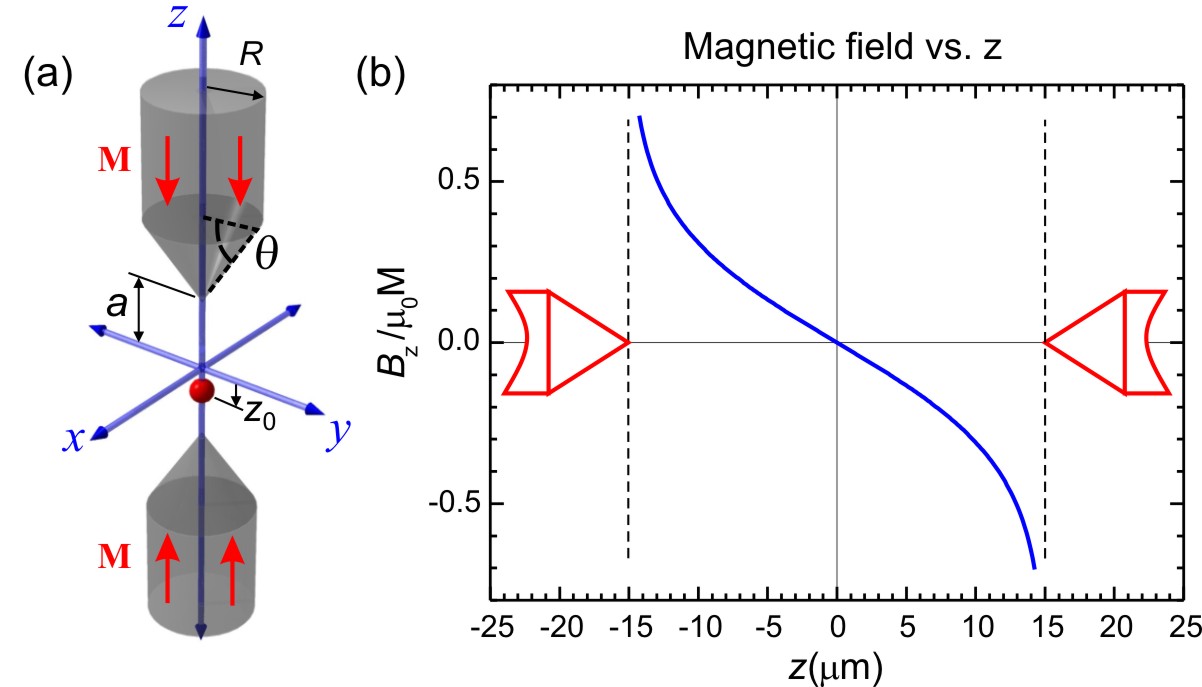}
	\caption{(a) Geometry of the magnetic tip trap with a trapped diamond near the center. (b) Field distribution along the $z$-axis for the "Reference magnetic trap" with $\theta =35\degree$(see text).}
	\label{fig:FigMTT}
\end{figure}

In the recent demonstration of a magnetic tip trap  O'Brien \textit{et~al.} uses  two cylindrical magnets with sharpened tips and a microdiamond as the trapped object \cite{ObrienAPL19}. The tips are separated by a gap $d=2a$ as shown in Fig.~\ref{fig:FigMTT}(a). The trapping occurs due to diamagnetic repulsion that balances the gravity of the diamond and the cylindrical symmetry that produces a stable potential confinement in three dimension. The study in Ref.~\cite{ObrienAPL19} has reported many important physical characteristics of the trap such as the vertical and radial trap frequency, damping factor and maximum field confinement at certain facet angle. However the detailed field and potential distribution of the magnetic trap has not been presented.

In this report we present a theoretical model that provides analytical solution of the magnetic field of the trap along its principal axis ($z$). The model leads to rich description such as equilibrium height, axial and radial oscillation frequency, the optimum facet angle of the tip to achieve the "confinement enhancement" and the critical gap beyond which such effect no longer applies. Beyond the recent interest in various magnetic traps, this model also serves as a new elementary example of a simple magnetic trap system based on conical tip geometry. This adds to the collection of various type of diamagnetic trap systems that have been known in physics \cite{BerryEJP97,SimonAJP01, LyuksyutovAPL04, GunawanAPL15, GunawanJAP17, HsuSR16,HoultonRSI18}.

We present a theoretical model of a pair of magnetic tips system as shown in Fig.~\ref{fig:FigMTT}(a). Each magnetic tip consists of a cylindrical segment of semi-infinite length and a conical segment. The magnet has a uniform volume magnetization ${\bf M}$ parallel to the cylindrical axis, however the two tips have opposing magnetization. The cylindrical segment has a radius $R$ and the conical tip has a facet angle $\theta$ \footnote{The experiment in Ref.~\cite{ObrienAPL19} uses magnetic tip with four flat facets. We have also modeled such system and found that our conical tip model is a very good approximation.}. The magnetic trap has a gap opening $d=2a$. A diamagnetic object such as a diamond bead can be trapped or levitates near the center of the trap at equilibrium position $z_0$.

We first consider the magnetic field along the principal axis $z$ due to the upper magnetic tip. We can calculate the magnetic field by integrating the field contributions due to bound surface current ${\bf K_b} = {\bf M} \times {\bf \hat{n}}$ all around the conical and the cylindrical segments, where ${\bf \hat{n}}$ is the normal of the surface element (see Supplementary Material (SM) A for detailed calculations). Interestingly this leads to a closed-form solution. The magnetic field due to the upper magnetic tip is given as:

\begin{widetext}
\begin{eqnarray}
{\bf B_1}(z) &=& - \frac{\mu_0 M}{2} \left[ \cos^2\theta [1 - \sin\theta ~\mathrm{arctanh}(\sin\theta)] + \cos^2\theta \sin\theta ~\mathrm{arctanh} \left( \frac{R + (a-z) \sin\theta \cos\theta}{\sqrt{R^2 + (a-z)^2 \cos^2\theta + R(a-z) \sin 2\theta}} \right) \right. \nonumber\\
&& \left. + \frac{z-a-R \tan\theta}{\sqrt{R^2 + (a-z+ R \tan\theta)^2}} + \frac{(a-z) \sin^2\theta \cos\theta - R \cos 2\theta \sin\theta}{\sqrt{R^2 + (a-z)^2 \cos^2\theta + R(a-z) \sin 2\theta}} \right] {\bf \hat{z}}.
\end{eqnarray}
\end{widetext}

By exploiting the symmetry of the problem, the total magnetic field due to both tips is given as: $B_T(z) = B_1(z) - B_1(-z)$. An example of total magnetic field plot along the $z$ axis (for $R = 1 ~\mathrm{mm}$ and $a = 15 ~\mathrm{\mu m}$ and $\theta = 35\degree$) is given in Fig.~\ref{fig:FigMTT}(b). We observe that the field distribution near the center of the trap is approximately linear which leads to a harmonic potential trap.  The total  potential of the trapped object  due to both magnetic interaction and the gravity per unit volume is given as:

\begin{equation}
U_T(z) = -\frac{\chi}{2 \mu_0} {B_T}^2(z) + \rho g z,
\end{equation}

\noindent where $\rho$ is the density of the trapped object, $\chi$ is the magnetic susceptibility, $g$ is the gravitational acceleration, and $\mu_0$ is the magnetic permeability in vacuum. We note that for a spherical diamagnetic object we should replace $\chi/2$ with $\chi/(2+\chi)$ \cite{ZangwillBook13}, however the former is a good approximation for very small $\chi$ as in the case of many diamagnetic materials (with exception of superconductor where $\chi=-1$). For the magnetic tip trap that we use in this study \cite{ObrienAPL19}, the magnet is made of NdFeB with volume magnetization $M = 10^6 ~\mathrm{A/m}$, and microdiamond as the trapped object with $\chi = -2.2 \times 10^{-5}$ \cite{HeremansPRB94} and $\rho = 3513 ~\mathrm{kg/m^3}$  \cite{DeanBook99}, $R = 1 ~\mathrm{mm}$ and $a = 15 ~\mathrm{\mu m}$. We refer to this setup as the "Reference magnetic trap" in this study. For further analysis, we define the feature size of the magnetic trap which is given as $\lambda_0 = |\chi| \mu_0 M^2/ \rho g$ which indicates the "strength" of the magnetic trap.  For the reference magnetic trap here we have $\lambda_0 = 805 ~\mathrm{\mu m}$.

%\begin{figure*}
%\includegraphics[width=0.7\textwidth]{Fig02.jpg}
%\caption{(Color online) (a) The equilibrium height (in terms of $z_0/a$) dependence on the half-gap (in terms of $a/R$) and the facet angle ($\theta$) for a magnetic trap with $\lambda_0 = 805 ~\mathrm{\mu m}$. The white dashed curve is the theoretical optimum facet angle (Eq.~3) and the black circles are the maximum point for $z_0/a$. The star is the data point for the Reference magnetic trap.(b) The trap frequency ($f_z$) dependence with respect to half-gap ($a/R$) and facet angle ($\theta$) for $\lambda_0 = 805 ~\mathrm{\mu m}$. (c) The equilibrium height ($z_0/a$) vs half-gap ($a/R$) with various magnetic trap strength ($\lambda_0$). (d) The critical half-gap ($a_c/R$) beyond which the confinement enhancement no longer applies, plotted with respect to the magnetic trap feature length $\lambda_0$ (at $\theta_{max}$=35.3\degree). The theory (Eq.~\ref{Qmax}) fits well for $\lambda_0/R < 2$.}
%\label{fig:fig:FigMTTPlot}
%\end{figure*}

\begin{figure*}
	\floatbox[{\capbeside\thisfloatsetup{capbesideposition={right,top},capbesidewidth=5cm}}]{figure}[\FBwidth]
	{\caption{(a) The equilibrium height ($z_0$) dependence on the half-gap ($a$) and the facet angle ($\theta$) for magnetic trap with  "strength" $\lambda_0=805 ~\mathrm{\mu m}$. The white dashed curve is the theoretical optimum facet angle (Eq.~\ref{eq:Qmax}) and the black circles are the maximum point for $z_0/a$. The star is the data point for the "Reference magnetic trap". (b) The trap frequency ($f_z$) dependence with respect to half-gap ($a$) and facet angle ($\theta$) for $\lambda_0=805 ~\mathrm{\mu m}$. (c) The equilibrium height ($z_0$) vs. half-gap ($a$) with various magnetic trap strength ($\lambda_0=10, 1,$ and $0.1 R$). (d) The critical half-gap ($a_c$) beyond which the confinement enhancement no longer applies, plotted with respect to the magnetic trap feature length $\lambda_0$ (at $\theta_{max}$=35.3\degree). The theory (Eq.~\ref{eq:a_c}) fits well for $\lambda_0/R < 2$.}\label{fig:FigMTTPlot}}
	{\includegraphics[width=12cm]{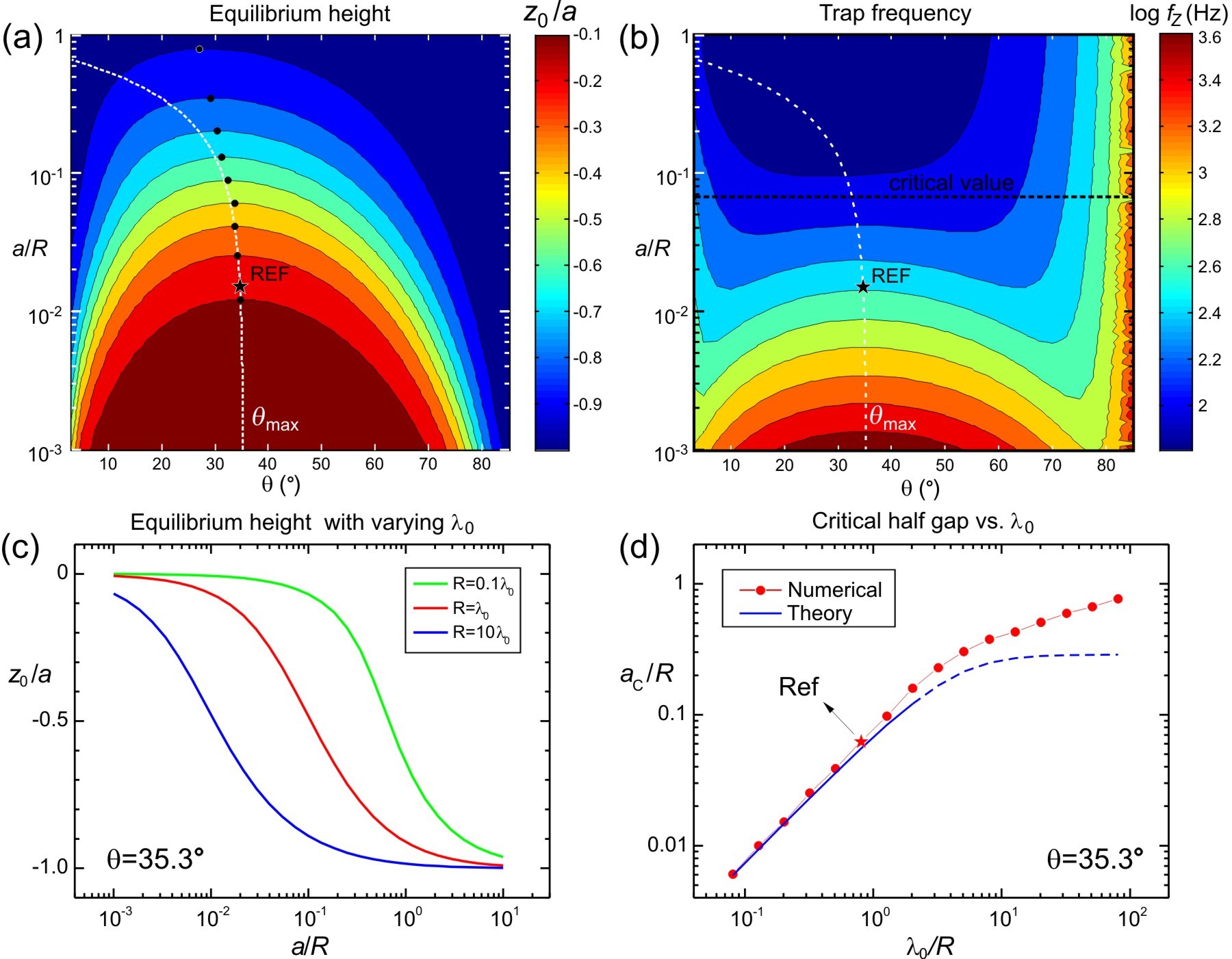}}
\end{figure*}

We now perform the analysis in the limit of a very strong trap (i.e. $\lambda_0 \gg a$) where the diamagnetic object will be trapped near the center ($z_0\sim 0$). As the magnetic field is reasonably linear at the center of the trap [Fig.~\ref{fig:FigMTT}(b)], the potential can be well approximated by harmonic potential model $U_T(z) \approx k_z z^2/2$, where $k_z$ is the "spring constant" given as: $k_z=\partial^2U_T/\partial z^2$.  We could obtain a very compact expression for the potential "spring constant" $k_z$ (per unit volume of the trapped object) in the strong magnetic trap limit where $z_0\sim 0$:

\begin{equation}
k_z = - \frac{\chi \mu_0 M^2 R^4 \cos^6\theta (a + R \tan\theta)^2}{a^2 (R^2 + a^2 \cos^2\theta + a R \sin 2\theta)^3}.
\label{kz}
\end{equation}

We show that this theoretical model provides rich descriptions of the magnetic trap characteristics that yields reasonable agreement with the experimental observation \cite{ObrienAPL19}. First, the model allows us to calculate the a natural frequency of the vertical oscillation of the trapped object: $f_z = \sqrt{k_z / \rho}/2\pi$ , which yields $f_z = 348 ~\mathrm{Hz}$. This is within the range of the reported frequency of $f_z = 323 ~\mathrm{Hz}$  to $411 ~\mathrm{Hz}$, at pressure of 760 Torr and 0.16 Torr respectively \cite{ObrienAPL19}. Note that the observed trapped frequency at room pressure (760 Torr) is lower due to significant damping effect.

Second, we can calculate the equilibrium position of the trapped object that yields $z_0 = -\rho g/k_z = -2.1 ~\mathrm{\mu m}$, which is small compared to the gap $d = 30 ~\mathrm{\mu m}$, in other words the object (microdiamond) remains near the center of the trap as observed \cite{ObrienAPL19}. We note that, theoretically, stable levitation always exists irrespective of the gap due to the magnetic field characteristics that diverges near the tip as shown in Fig.~\ref{fig:FigMTT}(b). When the gap is large, the object will levitate lower until it is balanced by the diamagnetic repulsion force which is proportional to $B ~dB/dz$.  However due to the finite size, the diamagnetic object will eventually touch the lower tip when the gap is large.

Third, the spring constant $k_z$ or the vertical trap frequency $f_z$ depend on the half gap separation $a$, the magnet cylindrical radius $R$ and the facet angle $\theta$. For a given $a$ and $R$, maximum frequency can be achieved at an optimum facet angle $\theta_{max}$. We can calculate this optimum facet angle which is given as (see SM B):
\begin{equation}
\theta_{max} = \arccos \sqrt{\frac{2R^2(2a^2 + 2\sqrt{2}aR + 3R^2)} {4a^4 + 4a^2 R^2 + 9R^4}}
\label{eq:Qmax}
\end{equation}
We note that in the strong magnetic trap limit ($\lambda_0>a$) the optimum facet angle depends only by geometrical factor of the magnetic tip i.e. $a/R$ and none of the physical properties of the magnet or the trapped object ($M, \chi, \rho$). For the "Reference magnetic trap" here, we obtain $\theta_{max} = 34.7^o$ which is quite close to reported value based on numerical calculation of boundary integral method for flat facets tips $\theta_{max} = 28\degree$ \cite{ObrienAPL19}. In the limit of very small gap ($a \ll R, \lambda_0$), Eq.~\ref{eq:Qmax} reduces to a very simple expression: $\theta_{max} = \arccos{\sqrt{2/3}} = 35.3\degree$.

Fourth, the experimental study also reported frequency of horizontal oscillation mode which is half that of vertical frequency i.e. $f_x = f_y = f_z/2 = 198 ~\mathrm{Hz}$ \cite{ObrienAPL19}. We note that for a magnetic trap with cylindrical symmetry such relationship is guaranteed i.e. $f_r = f_z/2$. The reason is, due to the fundamental characteristics of the static magnetic field, which are  divergence-free ($\nabla\cdot{\bf B}=0$) and curl-free ($\nabla\times{\bf B}=0$) we could relate the linear coefficients of the axial and radial magnetic field (\cite{McdonaldPU17}, pg. 4) as: $B_z(r,z) = a_0 + a_1(z-z_0) + ...$ and $B_r(r,z) = b_1 r + b_2 r(z-z_0) + ...$. This yields $b_1 = a_1/2$. Therefore, we have $f_r/f_z = \sqrt{k_r/k_z} = \sqrt{(b_1/a_1)^2} = 1/2$. This relationship applies to any magnetic trap with cylindrical symmetry (see SM C).

We now extend our analysis to the impact of wide range of geometrical and physical parameters ($a, R, \theta, \chi, \rho, M$) to the trap characteristics. We perform numerical calculation to obtain the vertical trap frequency which represents the strength of the trap confinement. First we start with the Reference magnetic trap ($\lambda_0 = 805 ~\mathrm{\mu m}$) and calculate the equilibrium position ($z_0$) of the trapped object in the magnetic trap by numerically solve for $dU_T/dz = 0$, and then we calculate the spring constant $k_z$ at that position ($z_0$). The equilibrium position $z_0$ as a function of half-gap $a$ and facet angle $\theta$ is plotted in Fig.~\ref{fig:FigMTTPlot}(a). We observe that for a constant $a$ value, we achieve maximum $z_0$ at $\theta \approx 35\degree$ (black circles). This maximum behavior apparently applies to all values of $a/R$ which implies that the magnetic trap always yields a higher levitation position $z_0$ at an optimum facet angle $\theta_{max}$ even when the gap is large. Next we study $z_0$ behavior with respect to half-gap $a$ at varying strength of magnetic trap ($\lambda_0$) as shown in Fig.~\ref{fig:FigMTTPlot}(c). We observe that for a very strong magnetic trap (e.g. where $\lambda_0 = 10R$) $z_0$ becomes higher than the others ($\lambda_0 = 1$ and $0.1 R$). It is near zero (near the center of the trap) until it drops off at higher gap value at $a/R>0.1$.

Next, after obtaining $z_0$ in Fig.~\ref{fig:FigMTTPlot}(a) we now numerically calculate the trap frequency $f_z = \sqrt{k_z(z_0)/\rho}/2\pi$ as a function of  $a$, $\theta$ and $z_0$ as plotted in Fig.~\ref{fig:FigMTTPlot}(b). First we observe a behavior similar to Fig.~\ref{fig:FigMTTPlot}(a) for small gap ($a/R < 0.07$), i.e. the magnetic trap achieves a maximum frequency at $\theta_{max}$ due to the confinement enhancement effect of the magnetic tip. This plot resembles the plot from numerical computation in Fig.~1(b) of Ref.~\cite{ObrienAPL19}. However we also observe an interesting behavior, for a large gap beyond a "critical value" $a_c$ this confinement enhancement effect no longer applies, i.e. when we plot $f_z$ vs. $\theta$ at constant $a$ - there is now a minimum near $\theta_{max}$. For the magnetic trap with very large gap ($a>a_c$), to achieve high frequency one can use very low facet angle (no tip) or very high (very sharp tip) which are not desirable. In practice we want to use smaller gap to achieve higher frequency, but not too small to provide some space for the trapped object and to allow optical detection.

Finally we study the dependence of this critical half-gap $a_c$ with respect to varying strength of magnetic trap by repeating the analysis in Fig.~\ref{fig:FigMTTPlot}(a) and (b) at different value of $\lambda_0$. The results is shown in Fig.~\ref{fig:FigMTTPlot}(d). We calculate $a_c$ using numerical calculations by finding the value $a$ where $\partial^2f_z/\partial\theta^2 = 0$ whose data are shown as red points in Fig.~\ref{fig:FigMTTPlot}(d). We observe a reasonable trend that the critical half-gap $a_c$ increases monotonically with increasing strength of the magnetic trap (or $\lambda_0$).

Furthermore we have also attempted to derive the theoretical relationship of this critical gap as a function of $\lambda_0$ using series expansion of $k_z$ as a function of $\lambda_0$ up to the 4th order (see SM D). We arrive at a simple relationship:

\begin{equation}
\frac{a_c}{R} = \frac{\lambda_0 \left( \sqrt{129.9R^2 + 3.74 \lambda_0^2} - \lambda_0 \right)}{153.86 R^2 + 3.24 \lambda_0^2}
\label{eq:a_c}
\end{equation}

\noindent This relationship allows us to quickly estimate the critical gap below which the confinement enhancement effect still applies in this magnetic tip trap. However, we also note that for strong magnetic trap ($\lambda_0/R > 2$) discrepancy occurs between the theoretical model and the numerical calculation due to higher order of $a$ that must be considered. We also note that for practical purpose, the optimum facet angle $\theta_{max} = \arccos{\sqrt{2/3}}$, which is calculated in the limit of strong trap, applies very well to most situation where $a < a_c$.

In closing, compared to other existing electromagnetic trap systems known in physics \cite{PenningPhys36, BrownRMP86, ChuPRL86, AshkinJQE00, ChuPRL85, RaabPRL87, PhillipsRMP98, SimonAJP01,LyuksyutovAPL04, GunawanAPL15, HsuSR16, HoultonRSI18}, this magnetic tip system presents a new, simple, elementary type of magnetic trap based on conical geometry. The model that we have developed provides rich theoretical understanding of the system that will help advance further development and applications.

\bibliography{PDL}

%merlin.mbs apsrev4-1.bst 2010-07-25 4.21a (PWD, AO, DPC) hacked
%Control: key (0)
%Control: author (8) initials jnrlst
%Control: editor formatted (1) identically to author
%Control: production of article title (-1) disabled
%Control: page (0) single
%Control: year (1) truncated
%Control: production of eprint (0) enabled
\begin{thebibliography}{32}%
\makeatletter
\providecommand \@ifxundefined [1]{%
 \@ifx{#1\undefined}
}%
\providecommand \@ifnum [1]{%
 \ifnum #1\expandafter \@firstoftwo
 \else \expandafter \@secondoftwo
 \fi
}%
\providecommand \@ifx [1]{%
 \ifx #1\expandafter \@firstoftwo
 \else \expandafter \@secondoftwo
 \fi
}%
\providecommand \natexlab [1]{#1}%
\providecommand \enquote  [1]{``#1''}%
\providecommand \bibnamefont  [1]{#1}%
\providecommand \bibfnamefont [1]{#1}%
\providecommand \citenamefont [1]{#1}%
\providecommand \href@noop [0]{\@secondoftwo}%
\providecommand \href [0]{\begingroup \@sanitize@url \@href}%
\providecommand \@href[1]{\@@startlink{#1}\@@href}%
\providecommand \@@href[1]{\endgroup#1\@@endlink}%
\providecommand \@sanitize@url [0]{\catcode `\\12\catcode `\$12\catcode
  `\&12\catcode `\#12\catcode `\^12\catcode `\_12\catcode `\%12\relax}%
\providecommand \@@startlink[1]{}%
\providecommand \@@endlink[0]{}%
\providecommand \url  [0]{\begingroup\@sanitize@url \@url }%
\providecommand \@url [1]{\endgroup\@href {#1}{\urlprefix }}%
\providecommand \urlprefix  [0]{URL }%
\providecommand \Eprint [0]{\href }%
\providecommand \doibase [0]{http://dx.doi.org/}%
\providecommand \selectlanguage [0]{\@gobble}%
\providecommand \bibinfo  [0]{\@secondoftwo}%
\providecommand \bibfield  [0]{\@secondoftwo}%
\providecommand \translation [1]{[#1]}%
\providecommand \BibitemOpen [0]{}%
\providecommand \bibitemStop [0]{}%
\providecommand \bibitemNoStop [0]{.\EOS\space}%
\providecommand \EOS [0]{\spacefactor3000\relax}%
\providecommand \BibitemShut  [1]{\csname bibitem#1\endcsname}%
\let\auto@bib@innerbib\@empty
%</preamble>
\bibitem [{\citenamefont {Penning}(1936)}]{PenningPhys36}%
  \BibitemOpen
  \bibfield  {author} {\bibinfo {author} {\bibfnamefont {F.~M.}\ \bibnamefont
  {Penning}},\ }\href@noop {} {\bibfield  {journal} {\bibinfo  {journal}
  {Physica (Utrecht)}\ }\textbf {\bibinfo {volume} {3}},\ \bibinfo {pages}
  {873} (\bibinfo {year} {1936})}\BibitemShut {NoStop}%
\bibitem [{\citenamefont {Brown}\ and\ \citenamefont
  {Gabrielse}(1986)}]{BrownRMP86}%
  \BibitemOpen
  \bibfield  {author} {\bibinfo {author} {\bibfnamefont {L.~S.}\ \bibnamefont
  {Brown}}\ and\ \bibinfo {author} {\bibfnamefont {G.}~\bibnamefont
  {Gabrielse}},\ }\href@noop {} {\bibfield  {journal} {\bibinfo  {journal}
  {Rev. Mod. Phys.}\ }\textbf {\bibinfo {volume} {58}},\ \bibinfo {pages} {233}
  (\bibinfo {year} {1986})}\BibitemShut {NoStop}%
\bibitem [{\citenamefont {Ashkin}(2000)}]{AshkinJQE00}%
  \BibitemOpen
  \bibfield  {author} {\bibinfo {author} {\bibfnamefont {A.}~\bibnamefont
  {Ashkin}},\ }\href@noop {} {\bibfield  {journal} {\bibinfo  {journal} {IEEE
  J. Quant. Elect.}\ }\textbf {\bibinfo {volume} {6}},\ \bibinfo {pages} {841}
  (\bibinfo {year} {2000})}\BibitemShut {NoStop}%
\bibitem [{\citenamefont {Chu}\ \emph {et~al.}(1986)\citenamefont {Chu},
  \citenamefont {Bjorkholm}, \citenamefont {Ashkin},\ and\ \citenamefont
  {Cable}}]{ChuPRL86}%
  \BibitemOpen
  \bibfield  {author} {\bibinfo {author} {\bibfnamefont {S.}~\bibnamefont
  {Chu}}, \bibinfo {author} {\bibfnamefont {J.}~\bibnamefont {Bjorkholm}},
  \bibinfo {author} {\bibfnamefont {A.}~\bibnamefont {Ashkin}}, \ and\ \bibinfo
  {author} {\bibfnamefont {A.}~\bibnamefont {Cable}},\ }\href@noop {}
  {\bibfield  {journal} {\bibinfo  {journal} {Phys. Rev. Lett.}\ }\textbf
  {\bibinfo {volume} {57}},\ \bibinfo {pages} {314} (\bibinfo {year}
  {1986})}\BibitemShut {NoStop}%
\bibitem [{\citenamefont {Chu}\ \emph {et~al.}(1985)\citenamefont {Chu},
  \citenamefont {Hollberg}, \citenamefont {Bjorkholm}, \citenamefont {Cable},\
  and\ \citenamefont {Ashkin}}]{ChuPRL85}%
  \BibitemOpen
  \bibfield  {author} {\bibinfo {author} {\bibfnamefont {S.}~\bibnamefont
  {Chu}}, \bibinfo {author} {\bibfnamefont {L.}~\bibnamefont {Hollberg}},
  \bibinfo {author} {\bibfnamefont {J.~E.}\ \bibnamefont {Bjorkholm}}, \bibinfo
  {author} {\bibfnamefont {A.}~\bibnamefont {Cable}}, \ and\ \bibinfo {author}
  {\bibfnamefont {A.}~\bibnamefont {Ashkin}},\ }\href {\doibase
  10.1103/PhysRevLett.55.48} {\bibfield  {journal} {\bibinfo  {journal} {Phys.
  Rev. Lett.}\ }\textbf {\bibinfo {volume} {55}},\ \bibinfo {pages} {48}
  (\bibinfo {year} {1985})}\BibitemShut {NoStop}%
\bibitem [{\citenamefont {Raab}\ \emph {et~al.}(1987)\citenamefont {Raab},
  \citenamefont {Prentiss}, \citenamefont {Cable}, \citenamefont {Chu},\ and\
  \citenamefont {Pritchard}}]{RaabPRL87}%
  \BibitemOpen
  \bibfield  {author} {\bibinfo {author} {\bibfnamefont {E.~L.}\ \bibnamefont
  {Raab}}, \bibinfo {author} {\bibfnamefont {M.}~\bibnamefont {Prentiss}},
  \bibinfo {author} {\bibfnamefont {A.}~\bibnamefont {Cable}}, \bibinfo
  {author} {\bibfnamefont {S.}~\bibnamefont {Chu}}, \ and\ \bibinfo {author}
  {\bibfnamefont {D.~E.}\ \bibnamefont {Pritchard}},\ }\href {\doibase
  10.1103/PhysRevLett.59.2631} {\bibfield  {journal} {\bibinfo  {journal} {Phys
  Rev Lett}\ }\textbf {\bibinfo {volume} {59}},\ \bibinfo {pages} {2631}
  (\bibinfo {year} {1987})}\BibitemShut {NoStop}%
\bibitem [{\citenamefont {Phillips}(1998)}]{PhillipsRMP98}%
  \BibitemOpen
  \bibfield  {author} {\bibinfo {author} {\bibfnamefont {W.~D.}\ \bibnamefont
  {Phillips}},\ }\href@noop {} {\bibfield  {journal} {\bibinfo  {journal} {Rev.
  Mod. Phys.}\ }\textbf {\bibinfo {volume} {70}},\ \bibinfo {pages} {721}
  (\bibinfo {year} {1998})}\BibitemShut {NoStop}%
\bibitem [{\citenamefont {Simon}\ \emph {et~al.}(2001)\citenamefont {Simon},
  \citenamefont {Heflinger},\ and\ \citenamefont {Geim}}]{SimonAJP01}%
  \BibitemOpen
  \bibfield  {author} {\bibinfo {author} {\bibfnamefont {M.~D.}\ \bibnamefont
  {Simon}}, \bibinfo {author} {\bibfnamefont {L.~O.}\ \bibnamefont
  {Heflinger}}, \ and\ \bibinfo {author} {\bibfnamefont {A.~K.}\ \bibnamefont
  {Geim}},\ }\href@noop {} {\bibfield  {journal} {\bibinfo  {journal} {Am. J.
  Phys.}\ }\textbf {\bibinfo {volume} {69}},\ \bibinfo {pages} {702} (\bibinfo
  {year} {2001})}\BibitemShut {NoStop}%
\bibitem [{\citenamefont {Lyuksyutov}\ \emph {et~al.}(2004)\citenamefont
  {Lyuksyutov}, \citenamefont {Naugle},\ and\ \citenamefont
  {Rathnayaka}}]{LyuksyutovAPL04}%
  \BibitemOpen
  \bibfield  {author} {\bibinfo {author} {\bibfnamefont {I.~F.}\ \bibnamefont
  {Lyuksyutov}}, \bibinfo {author} {\bibfnamefont {D.~G.}\ \bibnamefont
  {Naugle}}, \ and\ \bibinfo {author} {\bibfnamefont {K.~D.~D.}\ \bibnamefont
  {Rathnayaka}},\ }\href@noop {} {\bibfield  {journal} {\bibinfo  {journal}
  {Appl. Phys. Lett.}\ }\textbf {\bibinfo {volume} {85}},\ \bibinfo {pages}
  {1817} (\bibinfo {year} {2004})}\BibitemShut {NoStop}%
\bibitem [{\citenamefont {Gunawan}\ \emph {et~al.}(2015)\citenamefont
  {Gunawan}, \citenamefont {Virgus},\ and\ \citenamefont
  {Fai~Tai}}]{GunawanAPL15}%
  \BibitemOpen
  \bibfield  {author} {\bibinfo {author} {\bibfnamefont {O.}~\bibnamefont
  {Gunawan}}, \bibinfo {author} {\bibfnamefont {Y.}~\bibnamefont {Virgus}}, \
  and\ \bibinfo {author} {\bibfnamefont {K.}~\bibnamefont {Fai~Tai}},\ }\href
  {\doibase http://dx.doi.org/10.1063/1.4907931} {\bibfield  {journal}
  {\bibinfo  {journal} {Appl. Phys. Lett.}\ }\textbf {\bibinfo {volume}
  {106}},\ \bibinfo {pages} {062407} (\bibinfo {year} {2015})}\BibitemShut
  {NoStop}%
\bibitem [{\citenamefont {Hsu}\ \emph {et~al.}(2016)\citenamefont {Hsu},
  \citenamefont {Ji}, \citenamefont {Lewandowski},\ and\ \citenamefont
  {D'Urso}}]{HsuSR16}%
  \BibitemOpen
  \bibfield  {author} {\bibinfo {author} {\bibfnamefont {J.~F.}\ \bibnamefont
  {Hsu}}, \bibinfo {author} {\bibfnamefont {P.}~\bibnamefont {Ji}}, \bibinfo
  {author} {\bibfnamefont {C.~W.}\ \bibnamefont {Lewandowski}}, \ and\ \bibinfo
  {author} {\bibfnamefont {B.}~\bibnamefont {D'Urso}},\ }\href {\doibase
  10.1038/srep30125} {\bibfield  {journal} {\bibinfo  {journal} {Sci. Rep.}\
  }\textbf {\bibinfo {volume} {6}},\ \bibinfo {pages} {30125} (\bibinfo {year}
  {2016})}\BibitemShut {NoStop}%
\bibitem [{\citenamefont {Houlton}\ \emph {et~al.}(2018)\citenamefont
  {Houlton}, \citenamefont {Chen}, \citenamefont {Brubaker}, \citenamefont
  {Bertness},\ and\ \citenamefont {Rogers}}]{HoultonRSI18}%
  \BibitemOpen
  \bibfield  {author} {\bibinfo {author} {\bibfnamefont {J.~P.}\ \bibnamefont
  {Houlton}}, \bibinfo {author} {\bibfnamefont {M.~L.}\ \bibnamefont {Chen}},
  \bibinfo {author} {\bibfnamefont {M.~D.}\ \bibnamefont {Brubaker}}, \bibinfo
  {author} {\bibfnamefont {K.~A.}\ \bibnamefont {Bertness}}, \ and\ \bibinfo
  {author} {\bibfnamefont {C.~T.}\ \bibnamefont {Rogers}},\ }\href {\doibase
  10.1063/1.5051667} {\bibfield  {journal} {\bibinfo  {journal} {Review of
  Scientific Instruments}\ }\textbf {\bibinfo {volume} {89}},\ \bibinfo {pages}
  {125107} (\bibinfo {year} {2018})}\BibitemShut {NoStop}%
\bibitem [{\citenamefont {Simon}\ and\ \citenamefont
  {Geim}(2000)}]{SimonJAP00}%
  \BibitemOpen
  \bibfield  {author} {\bibinfo {author} {\bibfnamefont {M.~D.}\ \bibnamefont
  {Simon}}\ and\ \bibinfo {author} {\bibfnamefont {A.~K.}\ \bibnamefont
  {Geim}},\ }\href@noop {} {\bibfield  {journal} {\bibinfo  {journal} {J. Appl.
  Phys.}\ }\textbf {\bibinfo {volume} {87}},\ \bibinfo {pages} {6200} (\bibinfo
  {year} {2000})}\BibitemShut {NoStop}%
\bibitem [{\citenamefont {O'Brien}\ \emph {et~al.}(2019)\citenamefont
  {O'Brien}, \citenamefont {Dunn}, \citenamefont {Downes},\ and\ \citenamefont
  {Twamley}}]{ObrienAPL19}%
  \BibitemOpen
  \bibfield  {author} {\bibinfo {author} {\bibfnamefont {M.}~\bibnamefont
  {O'Brien}}, \bibinfo {author} {\bibfnamefont {S.}~\bibnamefont {Dunn}},
  \bibinfo {author} {\bibfnamefont {J.}~\bibnamefont {Downes}}, \ and\ \bibinfo
  {author} {\bibfnamefont {J.}~\bibnamefont {Twamley}},\ }\href@noop {}
  {\bibfield  {journal} {\bibinfo  {journal} {Appl. Phys. Lett.}\ }\textbf
  {\bibinfo {volume} {114}},\ \bibinfo {pages} {053103} (\bibinfo {year}
  {2019})}\BibitemShut {NoStop}%
\bibitem [{\citenamefont {Slezak}\ \emph {et~al.}(2018)\citenamefont {Slezak},
  \citenamefont {Lewandowski}, \citenamefont {Hsu},\ and\ \citenamefont
  {D’Urso}}]{SlezakNJP18}%
  \BibitemOpen
  \bibfield  {author} {\bibinfo {author} {\bibfnamefont {B.~R.}\ \bibnamefont
  {Slezak}}, \bibinfo {author} {\bibfnamefont {C.~W.}\ \bibnamefont
  {Lewandowski}}, \bibinfo {author} {\bibfnamefont {J.-F.}\ \bibnamefont
  {Hsu}}, \ and\ \bibinfo {author} {\bibfnamefont {B.}~\bibnamefont
  {D’Urso}},\ }\href {\doibase 10.1088/1367-2630/aacac1} {\bibfield
  {journal} {\bibinfo  {journal} {New J. Phys.}\ }\textbf {\bibinfo {volume}
  {20}},\ \bibinfo {pages} {063028} (\bibinfo {year} {2018})}\BibitemShut
  {NoStop}%
\bibitem [{\citenamefont {Mamin}\ \emph {et~al.}(2007)\citenamefont {Mamin},
  \citenamefont {Poggio}, \citenamefont {Degen},\ and\ \citenamefont
  {Rugar}}]{MaminNatNano07}%
  \BibitemOpen
  \bibfield  {author} {\bibinfo {author} {\bibfnamefont {H.~J.}\ \bibnamefont
  {Mamin}}, \bibinfo {author} {\bibfnamefont {M.}~\bibnamefont {Poggio}},
  \bibinfo {author} {\bibfnamefont {C.~L.}\ \bibnamefont {Degen}}, \ and\
  \bibinfo {author} {\bibfnamefont {D.}~\bibnamefont {Rugar}},\ }\href
  {\doibase 10.1038/nnano.2007.105} {\bibfield  {journal} {\bibinfo  {journal}
  {Nat. Nanotech.}\ }\textbf {\bibinfo {volume} {2}},\ \bibinfo {pages} {301}
  (\bibinfo {year} {2007})}\BibitemShut {NoStop}%
\bibitem [{\citenamefont {Rugar}\ \emph {et~al.}(2004)\citenamefont {Rugar},
  \citenamefont {Budakian}, \citenamefont {Mamin},\ and\ \citenamefont
  {Chui}}]{RugarNat04}%
  \BibitemOpen
  \bibfield  {author} {\bibinfo {author} {\bibfnamefont {D.}~\bibnamefont
  {Rugar}}, \bibinfo {author} {\bibfnamefont {R.}~\bibnamefont {Budakian}},
  \bibinfo {author} {\bibfnamefont {H.}~\bibnamefont {Mamin}}, \ and\ \bibinfo
  {author} {\bibfnamefont {B.}~\bibnamefont {Chui}},\ }\href@noop {} {\bibfield
   {journal} {\bibinfo  {journal} {Nature}\ }\textbf {\bibinfo {volume}
  {430}},\ \bibinfo {pages} {329} (\bibinfo {year} {2004})}\BibitemShut
  {NoStop}%
\bibitem [{\citenamefont {Johnsson}\ \emph {et~al.}(2016)\citenamefont
  {Johnsson}, \citenamefont {Brennen},\ and\ \citenamefont
  {Twamley}}]{JohnssonSR16}%
  \BibitemOpen
  \bibfield  {author} {\bibinfo {author} {\bibfnamefont {M.~T.}\ \bibnamefont
  {Johnsson}}, \bibinfo {author} {\bibfnamefont {G.~K.}\ \bibnamefont
  {Brennen}}, \ and\ \bibinfo {author} {\bibfnamefont {J.}~\bibnamefont
  {Twamley}},\ }\href {\doibase 10.1038/srep37495} {\bibfield  {journal}
  {\bibinfo  {journal} {Sci. Rep.}\ }\textbf {\bibinfo {volume} {6}},\ \bibinfo
  {pages} {37495} (\bibinfo {year} {2016})}\BibitemShut {NoStop}%
\bibitem [{\citenamefont {Li}\ \emph {et~al.}(2010)\citenamefont {Li},
  \citenamefont {Kheifets}, \citenamefont {Medellin},\ and\ \citenamefont
  {Raizen}}]{LiSci10}%
  \BibitemOpen
  \bibfield  {author} {\bibinfo {author} {\bibfnamefont {T.}~\bibnamefont
  {Li}}, \bibinfo {author} {\bibfnamefont {S.}~\bibnamefont {Kheifets}},
  \bibinfo {author} {\bibfnamefont {D.}~\bibnamefont {Medellin}}, \ and\
  \bibinfo {author} {\bibfnamefont {M.~G.}\ \bibnamefont {Raizen}},\
  }\href@noop {} {\bibfield  {journal} {\bibinfo  {journal} {Science}\ }\textbf
  {\bibinfo {volume} {328}},\ \bibinfo {pages} {1673} (\bibinfo {year}
  {2010})}\BibitemShut {NoStop}%
\bibitem [{\citenamefont {Millen}\ \emph {et~al.}(2014)\citenamefont {Millen},
  \citenamefont {Deesuwan}, \citenamefont {Barker},\ and\ \citenamefont
  {Anders}}]{MillenNatNano14}%
  \BibitemOpen
  \bibfield  {author} {\bibinfo {author} {\bibfnamefont {J.}~\bibnamefont
  {Millen}}, \bibinfo {author} {\bibfnamefont {T.}~\bibnamefont {Deesuwan}},
  \bibinfo {author} {\bibfnamefont {P.}~\bibnamefont {Barker}}, \ and\ \bibinfo
  {author} {\bibfnamefont {J.}~\bibnamefont {Anders}},\ }\href {\doibase
  10.1038/nnano.2014.82} {\bibfield  {journal} {\bibinfo  {journal} {Nat.
  Nanotech.}\ }\textbf {\bibinfo {volume} {9}},\ \bibinfo {pages} {425}
  (\bibinfo {year} {2014})}\BibitemShut {NoStop}%
\bibitem [{\citenamefont {Yin}\ \emph {et~al.}(2013)\citenamefont {Yin},
  \citenamefont {Li}, \citenamefont {Zhang},\ and\ \citenamefont
  {Duan}}]{YinPRA13}%
  \BibitemOpen
  \bibfield  {author} {\bibinfo {author} {\bibfnamefont {Z.-q.}\ \bibnamefont
  {Yin}}, \bibinfo {author} {\bibfnamefont {T.}~\bibnamefont {Li}}, \bibinfo
  {author} {\bibfnamefont {X.}~\bibnamefont {Zhang}}, \ and\ \bibinfo {author}
  {\bibfnamefont {L.}~\bibnamefont {Duan}},\ }\href@noop {} {\bibfield
  {journal} {\bibinfo  {journal} {Phys. Rev. A}\ }\textbf {\bibinfo {volume}
  {88}},\ \bibinfo {pages} {033614} (\bibinfo {year} {2013})}\BibitemShut
  {NoStop}%
\bibitem [{\citenamefont {Scala}\ \emph {et~al.}(2013)\citenamefont {Scala},
  \citenamefont {Kim}, \citenamefont {Morley}, \citenamefont {Barker},\ and\
  \citenamefont {Bose}}]{ScalaPRL13}%
  \BibitemOpen
  \bibfield  {author} {\bibinfo {author} {\bibfnamefont {M.}~\bibnamefont
  {Scala}}, \bibinfo {author} {\bibfnamefont {M.~S.}\ \bibnamefont {Kim}},
  \bibinfo {author} {\bibfnamefont {G.~W.}\ \bibnamefont {Morley}}, \bibinfo
  {author} {\bibfnamefont {P.~F.}\ \bibnamefont {Barker}}, \ and\ \bibinfo
  {author} {\bibfnamefont {S.}~\bibnamefont {Bose}},\ }\href {\doibase
  10.1103/PhysRevLett.111.180403} {\bibfield  {journal} {\bibinfo  {journal}
  {Phys. Rev. Lett.}\ }\textbf {\bibinfo {volume} {111}},\ \bibinfo {pages}
  {180403} (\bibinfo {year} {2013})}\BibitemShut {NoStop}%
\bibitem [{\citenamefont {Hoang}\ \emph {et~al.}(2016)\citenamefont {Hoang},
  \citenamefont {Ahn}, \citenamefont {Bang},\ and\ \citenamefont
  {Li}}]{HoangNatComm16}%
  \BibitemOpen
  \bibfield  {author} {\bibinfo {author} {\bibfnamefont {T.~M.}\ \bibnamefont
  {Hoang}}, \bibinfo {author} {\bibfnamefont {J.}~\bibnamefont {Ahn}}, \bibinfo
  {author} {\bibfnamefont {J.}~\bibnamefont {Bang}}, \ and\ \bibinfo {author}
  {\bibfnamefont {T.}~\bibnamefont {Li}},\ }\href {\doibase
  10.1038/ncomms12250} {\bibfield  {journal} {\bibinfo  {journal} {Nat. Comm.}\
  }\textbf {\bibinfo {volume} {7}},\ \bibinfo {pages} {12250} (\bibinfo {year}
  {2016})}\BibitemShut {NoStop}%
\bibitem [{\citenamefont {Kafri}\ \emph {et~al.}(2014)\citenamefont {Kafri},
  \citenamefont {Taylor},\ and\ \citenamefont {Milburn}}]{KafriNJP14}%
  \BibitemOpen
  \bibfield  {author} {\bibinfo {author} {\bibfnamefont {D.}~\bibnamefont
  {Kafri}}, \bibinfo {author} {\bibfnamefont {J.~M.}\ \bibnamefont {Taylor}}, \
  and\ \bibinfo {author} {\bibfnamefont {G.~J.}\ \bibnamefont {Milburn}},\
  }\href {\doibase 10.1088/1367-2630/16/6/065020} {\bibfield  {journal}
  {\bibinfo  {journal} {New J. Phys.}\ }\textbf {\bibinfo {volume} {16}},\
  \bibinfo {pages} {065020} (\bibinfo {year} {2014})}\BibitemShut {NoStop}%
\bibitem [{\citenamefont {Bose}\ \emph {et~al.}(2017)\citenamefont {Bose},
  \citenamefont {Mazumdar}, \citenamefont {Morley}, \citenamefont {Ulbricht},
  \citenamefont {Toros}, \citenamefont {Paternostro}, \citenamefont {Geraci},
  \citenamefont {Barker}, \citenamefont {Kim},\ and\ \citenamefont
  {Milburn}}]{BosePRL17}%
  \BibitemOpen
  \bibfield  {author} {\bibinfo {author} {\bibfnamefont {S.}~\bibnamefont
  {Bose}}, \bibinfo {author} {\bibfnamefont {A.}~\bibnamefont {Mazumdar}},
  \bibinfo {author} {\bibfnamefont {G.~W.}\ \bibnamefont {Morley}}, \bibinfo
  {author} {\bibfnamefont {H.}~\bibnamefont {Ulbricht}}, \bibinfo {author}
  {\bibfnamefont {M.}~\bibnamefont {Toros}}, \bibinfo {author} {\bibfnamefont
  {M.}~\bibnamefont {Paternostro}}, \bibinfo {author} {\bibfnamefont {A.~A.}\
  \bibnamefont {Geraci}}, \bibinfo {author} {\bibfnamefont {P.~F.}\
  \bibnamefont {Barker}}, \bibinfo {author} {\bibfnamefont {M.~S.}\
  \bibnamefont {Kim}}, \ and\ \bibinfo {author} {\bibfnamefont
  {G.}~\bibnamefont {Milburn}},\ }\href {\doibase
  10.1103/PhysRevLett.119.240401} {\bibfield  {journal} {\bibinfo  {journal}
  {Phys. Rev. Lett.}\ }\textbf {\bibinfo {volume} {119}},\ \bibinfo {pages}
  {240401} (\bibinfo {year} {2017})}\BibitemShut {NoStop}%
\bibitem [{\citenamefont {Berry}\ and\ \citenamefont
  {Geim}(1997)}]{BerryEJP97}%
  \BibitemOpen
  \bibfield  {author} {\bibinfo {author} {\bibfnamefont {M.~V.}\ \bibnamefont
  {Berry}}\ and\ \bibinfo {author} {\bibfnamefont {A.~K.}\ \bibnamefont
  {Geim}},\ }\href@noop {} {\bibfield  {journal} {\bibinfo  {journal} {Eur. J.
  Phys.}\ }\textbf {\bibinfo {volume} {18}},\ \bibinfo {pages} {307} (\bibinfo
  {year} {1997})}\BibitemShut {NoStop}%
\bibitem [{\citenamefont {Gunawan}\ and\ \citenamefont
  {Virgus}(2017)}]{GunawanJAP17}%
  \BibitemOpen
  \bibfield  {author} {\bibinfo {author} {\bibfnamefont {O.}~\bibnamefont
  {Gunawan}}\ and\ \bibinfo {author} {\bibfnamefont {Y.}~\bibnamefont
  {Virgus}},\ }\href {\doibase 10.1063/1.4978876]} {\bibfield  {journal}
  {\bibinfo  {journal} {J. Appl. Phys.}\ }\textbf {\bibinfo {volume} {121}},\
  \bibinfo {pages} {133902} (\bibinfo {year} {2017})}\BibitemShut {NoStop}%
\bibitem [{Note1()}]{Note1}%
  \BibitemOpen
  \bibinfo {note} {The experiment in Ref.~\cite {ObrienAPL19} uses magnetic tip
  with four flat facets. We have also modeled such system and found that our
  conical tip model is a very good approximation.}\BibitemShut {Stop}%
\bibitem [{\citenamefont {Zangwill}(2013)}]{ZangwillBook13}%
  \BibitemOpen
  \bibfield  {author} {\bibinfo {author} {\bibfnamefont {A.}~\bibnamefont
  {Zangwill}},\ }\href@noop {} {\emph {\bibinfo {title} {Modern
  electrodynamics}}}\ (\bibinfo  {publisher} {Cambridge University Press},\
  \bibinfo {year} {2013})\BibitemShut {NoStop}%
\bibitem [{\citenamefont {Heremans}\ \emph {et~al.}(1994)\citenamefont
  {Heremans}, \citenamefont {Olk},\ and\ \citenamefont
  {Morelli}}]{HeremansPRB94}%
  \BibitemOpen
  \bibfield  {author} {\bibinfo {author} {\bibfnamefont {J.}~\bibnamefont
  {Heremans}}, \bibinfo {author} {\bibfnamefont {C.~H.}\ \bibnamefont {Olk}}, \
  and\ \bibinfo {author} {\bibfnamefont {D.~T.}\ \bibnamefont {Morelli}},\
  }\href {\doibase 10.1103/PhysRevB.49.15122} {\bibfield  {journal} {\bibinfo
  {journal} {Phys. Rev. B}\ }\textbf {\bibinfo {volume} {49}},\ \bibinfo
  {pages} {15122} (\bibinfo {year} {1994})}\BibitemShut {NoStop}%
\bibitem [{\citenamefont {Dean}(1999)}]{DeanBook99}%
  \BibitemOpen
  \bibfield  {author} {\bibinfo {author} {\bibfnamefont {J.~A.}\ \bibnamefont
  {Dean}},\ }\href@noop {} {\emph {\bibinfo {title} {Lange's handbook of
  chemistry}}}\ (\bibinfo  {publisher} {New york; London: McGraw-Hill, Inc.},\
  \bibinfo {year} {1999})\BibitemShut {NoStop}%
\bibitem [{\citenamefont {McDonald}(2015)}]{McdonaldPU17}%
  \BibitemOpen
  \bibfield  {author} {\bibinfo {author} {\bibfnamefont {K.~T.}\ \bibnamefont
  {McDonald}},\ }\href@noop {} {\bibfield  {journal} {\bibinfo  {journal}
  {http://www.hep.princeton.edu/~mcdonald/examples/diamagnetic.pdf for
  “Diamagnetic levitation"}\ } (\bibinfo {year} {March 4, 2015})}\BibitemShut
  {NoStop}%
\end{thebibliography}%

\end{document}

\maketitle
\section*{A. Magnetic Field Calculation}
\begin{figure}[H]
\centering
\includegraphics[width=0.7\textwidth]{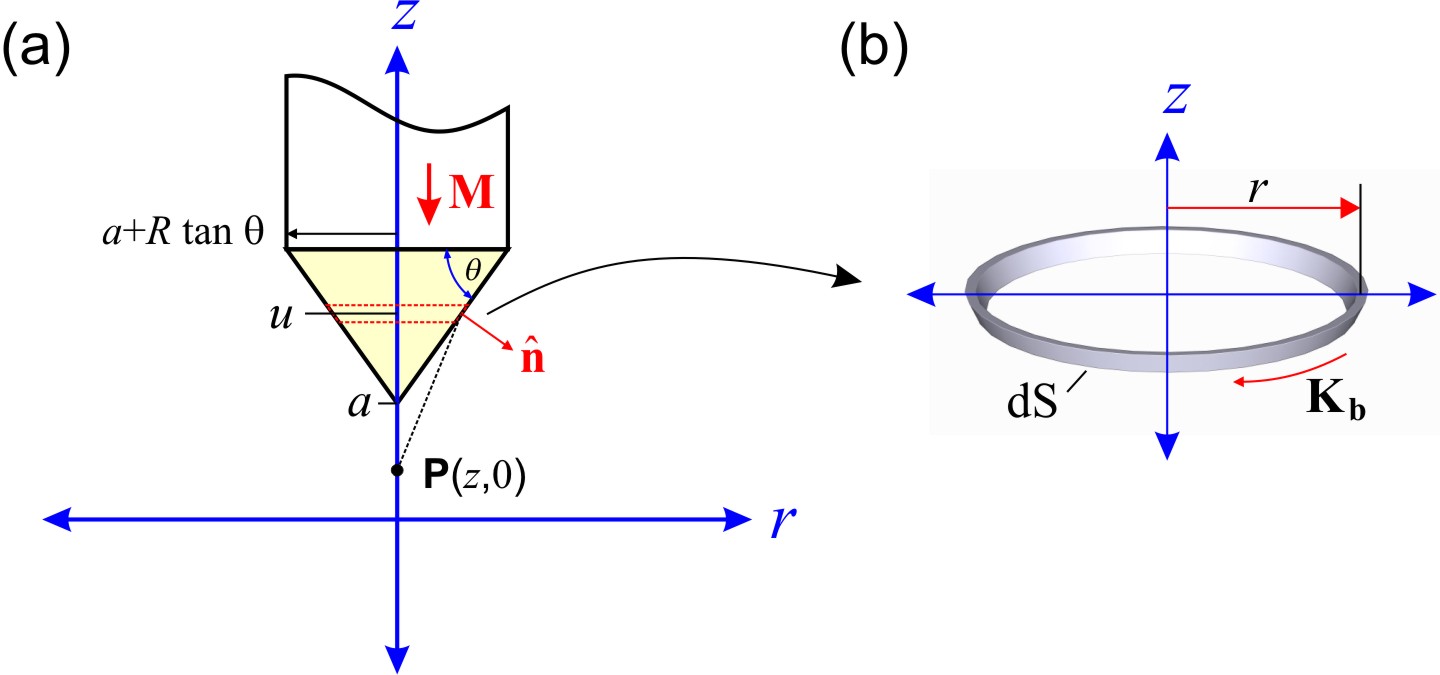}
\caption{(a) A magnetic tip trap model. (b) Current loop on the cone surface.}
\label{fig:FigS1}
\end{figure}

We will calculate the magnetic field along the principal axis ($z$) of a pair of magnetic tips as shown in Fig.~1(a) in the main text. First we consider a single magnetic tip as shown in Fig.~\ref{fig:FigS1}(a). The magnetic tip system can be modeled as a semi-infinite cylindrical section and a conical tip. The magnet has a uniform magnetization ${\bf M}$ with direction along the principal axis pointing toward the tip. The tip is located at $z=a$, the conic has radius $R$ and "facet angle" $\theta$ with respect to horizontal, as shown in Fig.~\ref{fig:FigS1}.
 
First consider the conical tip that has bound surface current element at $z=u$ due to the magnetization. The surface current forms a current loop with radius $r = (u-a) \cot\theta$, that produces magnetic field:

\begin{eqnarray}
d{\bf B_c} &=& \frac{\mu_0}{2} \frac{r^2 ~dI}{(r^2 + \Delta z^2)^{3/2}} (- {\bf \hat{z}}) \nonumber\\
&=& \frac{\mu_0}{2} \frac{(u-a)^2 \cot^2\theta ~dI}{[(u-a)^2 \cot^2\theta + (u-z)^2]^{3/2}} (- {\bf \hat{z}}).
\end{eqnarray}

The surface current element on the conical surface can be calculated as: ${\bf K_b}= {\bf M}\cross{\bf \hat{n}}$ where ${\bf \hat{n}}$ is the surface normal.  The current element is given as: $dI = K_b ~dS$, where $K = M \sin\theta$ and $dS = du/\sin\theta$. As a result, the current element is equal to $dI = M ~du$. The magnetic field produced by the conical tip is given by:

\begin{equation}
{\bf B_c}(z) = - \frac{\mu_0 M}{2} \int_a^{a+R \tan\theta} \frac{(u-a)^2 \cot^2\theta ~du}{[(u-a)^2 \cot^2\theta + (u-z)^2]^{3/2}}{\bf \hat{z}}.
\end{equation}

\noindent The integral yields analytical result:
\begin{eqnarray}
{\bf B_c}(z) &=& - \frac{\mu_0 M}{2} {\bf \hat{z}} \left[ \frac{(a-z) \sin^2\theta \cos\theta - R \cos 2\theta \sin\theta}{\sqrt{R^2 + (a-z)^2 \cos^2\theta + R(a-z) \sin 2\theta}} - \cos^2\theta \sin\theta ~\mathrm{arctanh}(\sin\theta) \right. \nonumber\\
&& \left. \cos^2\theta \sin\theta ~\mathrm{arctanh} \left( \frac{R + (a-z) \sin\theta \cos\theta}{\sqrt{R^2 + (a-z)^2 \cos^2\theta + R(a-z) \sin 2\theta}} \right) - \sin^2\theta \right].
\end{eqnarray}

\begin{figure}[H]
\centering
\includegraphics[width=0.4\textwidth]{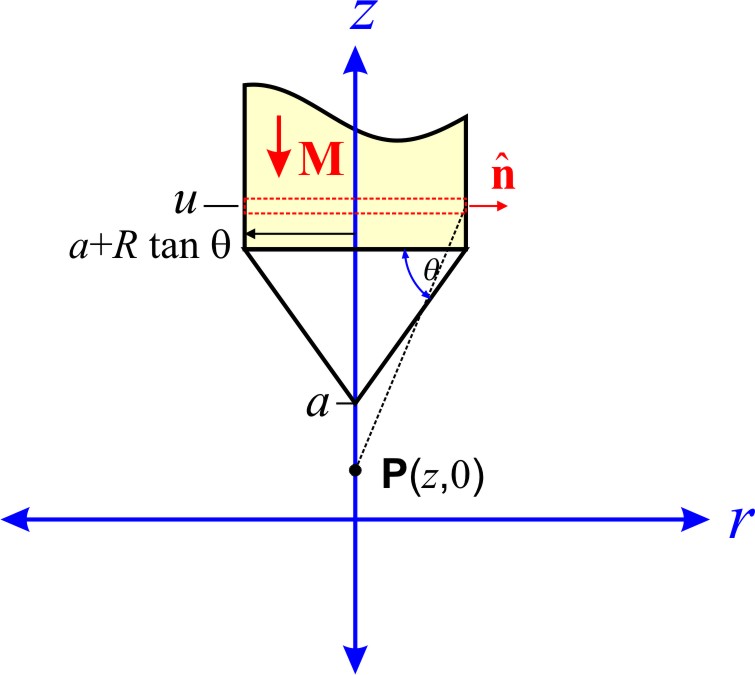}
\caption{Magnetic field produced by the cylindrical sheath.}
\label{fig:FigS2}
\end{figure}

Next we calculate the magnetic field due to the cylindrical sheath, which is straightforward. Consider a bound surface current element at $z = u$ with radius $R$ and current $dI = M ~du$, as in Fig. \ref{fig:FigS2}. The magnetic field produced by magnet sheath is given by:

\begin{eqnarray}
{\bf B_s}(z) &=& - \frac{\mu_0 M}{2} {\bf \hat{z}} \int_{a+R \tan\theta}^\infty \frac{R^2 ~du}{[R^2 + (u-z)^2]^{3/2}} \nonumber\\
&=& - \frac{\mu_0 M}{2} {\bf \hat{z}} \left[ 1 + \frac{z - a - R \tan\theta}{\sqrt{R^2 + (a + R \tan\theta -z)^2}} \right]
\end{eqnarray}

Total magnetic field due to the conical tip and the sheath is given as:

\begin{eqnarray}
{\bf B_1}(z) &=& - \frac{\mu_0 M}{2} {\bf \hat{z}} \left[ \cos^2\theta [1 - \sin\theta ~\mathrm{arctanh}(\sin\theta)] + \right. \nonumber\\
&& \left. \cos^2\theta \sin\theta ~\mathrm{arctanh} \left[ \frac{R + (a-z) \sin\theta \cos\theta}{\sqrt{R^2 + (a-z)^2 \cos^2\theta + R(a-z) \sin 2\theta}} \right] + \right. \nonumber\\
&& \left. \frac{z-a-R \tan\theta}{\sqrt{R^2 + (a-z+ R \tan\theta)^2}} + \frac{(a-z) \sin^2\theta \cos\theta - R \cos 2\theta \sin\theta}{\sqrt{R^2 + (a-z)^2 \cos^2\theta + R(a-z) \sin 2\theta}} \right].
\end{eqnarray}

Finally, by exploiting the symmetry of the problem, the total magnetic field of the upper and lower magnetic tip is given as:

\begin{equation}
{\bf B_T}(z) = {\bf B_1}(z) - {\bf B_1}(-z).
\label{eq:EqB_T}
\end{equation}

\section*{B. Trap Frequency and Optimum Facet Angle}

The total trap potential per unit volume of trapped-object along $z$ can be approximated as harmonic potential due to magnetic interaction plus gravitational term:

\begin{equation}
U_T(z) = \frac{1}{2}k_z z^2 + \rho g z
\end{equation}
where $k_z$ is the "spring constant" per unit volume: 
\begin{equation}
k_z = \frac{\partial^2 U_M}{\partial z^2}
\end{equation}
and $U_M$ is the magnetic energy potential per unit volume given as:
\begin{equation}
U_M(z) = -\frac{\chi B_T^2(z)}{2 \mu_0}. 
\end{equation}

\noindent Using $B_T$ calculated from Eq.~\ref{eq:EqB_T}, we can obtain $k_z$ in a surprisingly compact form:
\begin{equation}
k_z = - \frac{\chi \mu_0 M^2 R^4 \cos^6\theta (a + R \tan\theta)^2}{a^2 (R^2 + a^2 \cos^2\theta + a R \sin 2\theta)^3}.
\end{equation}

\noindent Condition $dk_z/d\theta = 0$ implies the optimum facet angle $\theta_{max}$ is obtained for $\theta$ that satisfies:
\begin{equation}
\frac{d}{d\theta} \left[ \frac{\cos^6 \theta (a+R \tan\theta)^2}{(R^2 + a^2 \cos^2 \theta + aR \sin 2\theta)^3} \right] = 0.
\end{equation}

\noindent By using $\eta = a/R$, the equation which is satisfied by $\theta_{max}$ is:
\begin{equation}
(9+4\eta^2+4\eta^4) \cos^4\theta_{max} - (12+8\eta^2) \cos^2\theta_{max} + 4 = 0.
\end{equation}

\noindent This yields solution for $\theta_{max}$ as:
\begin{equation}
\theta_{max} = \arccos \sqrt{\frac{2(3 + 2\eta^2 + 2\sqrt{2}\eta)} {9+4\eta^2+4\eta^4}}.
\end{equation}

In the limit of small gap $(a \ll R)$, the optimum facet angle reduces to a very simple expression:
\begin{equation}
\theta_{max} \approx \arccos \sqrt{2/3} = 35.3\degree.
\end{equation}

\section*{C. Magnetic Trap with Cylindrical Symmetry}
We will try to find relationship between the axial and radial frequency of an object trapped in a magnetic trap with cylindrical symmetry. First we consider the divergence-free property of the magnetic field: $\nabla \cdot {\bf B}=0$ or equivalently the Gauss Law for magnetic field $\oint{\bf B \cdot dA}=0$. For a magnetic field with cylindrical symmetry, the flux at the bottom of the cylinder is [32]:
\begin{equation}
- \int B_z (2 \pi r) ~dr
\end{equation}
and the flux at the top of the cylinder is:
\begin{equation}
\int (B_z + dB_z) (2\pi r) ~dr.
\end{equation}
At the cylindrical sheath, the magnetic flux is:
\begin{equation}
B_r (2\pi r) ~dz.
\end{equation}
Total of the three component should be zero, so we have:
\begin{equation}
- \int B_z (2 \pi r) ~dr + \int (B_z + dB_z) (2\pi r) ~dr + B_r (2\pi r) ~dz = 0
\end{equation}
\begin{equation}
B_r (r,z) = - \frac{1}{r} \int \frac{\partial B_z}{\partial z} r ~dr
\label{eq:Brad}
\end{equation}
A general form of $B_z$ near $z = z_0$ and $r = 0$ up to second order is given by:
\begin{equation}
B_z (r,z) = b_0 + b_1 (z-z_0) + b_2 (z-z_0)^2 + b_3 r + b_4 r^2 + b_5 (z-z_0)r.
\end{equation}
Now we can calculate \eqref{eq:Brad}:
\begin{equation}
B_r (r,z) = -\frac{1}{r} \left[ \frac{1}{2} b_1 r^2 + b_2 (z-z_0) r^2 + \frac{1}{3} b_5 r^3 \right] + c.
\end{equation}
Constant $c$ is zero because boundary condition $B_r(0,z = 0)$. Since $\nabla\cross{\bf B}=0$, both the axial and radial magnetic field must satisfy:
\begin{equation}
\frac{\partial B_r}{\partial z} - \frac{\partial B_z}{\partial r} = 0,
\end{equation}
\begin{equation}
-b_2 r - b_3 - 2 b_4 r - b_5 (z-z_0) = 0.
\end{equation}
To make the equation consistent for all values of $r$ and $z$, the coefficients must be $b_3 = 0$, $b_5 =0$, and $b_4 = -b_2 /2$. Finally, we get both the axial and radial magnetic field up to second order:
\begin{equation}
B_z (r,z) = b_0 + b_1 (z-z_0) + b_2 \left[ (z-z_0)^2 - \frac{r^2}{2} \right],
\end{equation}
\begin{equation}
B_r (r,z) = -\frac{1}{2}b_1 r - b_2 (z-z_0) r.
\end{equation}
Now we can calculate the ratio of the radial and the axial frequencies: $f_r/f_z= \sqrt{k_r/k_z}$. The spring constant for the axial component:
\begin{equation}
k_z=\partial^2U_T/\partial z^2 \propto \partial^2 B_z(r=0,z)^2/\partial z^2 \propto b_1^2
\end{equation}
The spring constant for the radial component:
\begin{equation}
k_r=\partial^2U_T/\partial r^2 \propto \partial^2 B_r(r,z=z_0)^2/\partial r^2 \propto (b_1/2)^2
\end{equation}
Therefore, we have: $f_r/f_z= (b_1/2)/b1 = 1/2$.

\section*{D. Critical Gap}
In order to find the critical gap, i.e. the gap beyond which there is no more confinement enhancement effect, we can expand the magnetic energy up to the fourth-order:
\begin{equation}
U_M(z) \approx - \frac{\chi}{8} \mu_0 M^2 (\alpha z^2 + \beta z^4)
\end{equation}
\begin{equation}
\alpha = \frac{32 R^4 \cos^4\theta (a \cos\theta + R \sin\theta)^2}{a^2 (a^2 + 2R^2 + a^2 \cos 2\theta + 2 a R \sin 2\theta)^3}
\end{equation}
\begin{eqnarray}
\beta &=& \frac{8R^4 \cos^4\theta (a \cos\theta + R \sin\theta)}{3a^4 (a^2 + 2R^2 + a^2 \cos 2\theta + 2 a R \sin 2\theta)^5} \left[ 2a (60a^4 + 47 a^2 R^2 + 28R^4) \cos\theta + \right. \nonumber\\
&& \left. a (60a^4 - 77 a^2 R^2 - 56R^4) \cos 3\theta + a^3 (12a^2 - 65R^2) \cos 5\theta + \right. \nonumber\\
&& \left. 2R (44a^4 + 49a^2 R^2 + 16R^4) \sin\theta + 3a^2 R (44a^2 + 21R^2) \sin 3\theta + \right. \nonumber\\
&& \left. a^2 R (44a^2 - 35R^2) \sin 5\theta \right] 
\end{eqnarray}
The equilibrium height of diamond $z_0$ is the solution of:
\begin{equation}
U_T(z_0) \approx -\frac{\chi}{4} \mu_0 M^2 \alpha z_0 + \rho g = 0,
\end{equation}
so we can express $z_0$ in terms of $\lambda_0 = |\chi| \mu_0 M^2 / \rho g$ and $\alpha$:
\begin{equation}
z_0 = \frac{4}{\lambda_0 \alpha}.
\end{equation}
The spring constant up to leading order of $z_0$ is:
\begin{eqnarray}
k_z(\theta, z_0) &=& - \frac{\chi}{4} \mu_0 M^2 (\alpha + 6 \beta {z_0}^2) \nonumber\\
&=& - \frac{\chi}{4} \mu_0 M^2 \left( \alpha + \frac{96\beta}{\lambda_0^2 \alpha^2} \right).
\end{eqnarray}
Confinement enhancement exists if there is an optimum angle $\theta_{max}$  where the spring constant $k_z$ becomes maximum. However there is a critical gap (or half-gap $a_c$) beyond which there is no $\theta$ between 0 and 90$\degree$ that yields maximum  $k_z$. The condition for the critical gap is: 
\begin{equation}
\frac{\partial^2 k_z}{\partial \theta^2} = 0,
\end{equation}
evaluated at optimum angle $\theta_{max} = 35.3\degree$. The relation between $a_c$, $R$, and $\lambda_0$ to satisfy that condition is:
\begin{equation}
-\frac{7.108}{a^2} + \frac{16.839}{aR} + \frac{1295.442(R^2+0.021 \lambda_0^2)}{R^2 \lambda_0^2} = 0.
\end{equation}
Hence, the solution for critical gap is:
\begin{equation}
\frac{a_c}{R} = \frac{\lambda_0 (\sqrt{129.9R^2 + 3.74 \lambda_0^2} - \lambda_0)}{153.86R^2 + 3.24 \lambda_0^2}.
\end{equation}